\documentstyle[12pt]{article}
\newcommand{\newsection}{    
\setcounter{equation}{0}
\section}

\def\eop{\vspace*{\fill}\pagebreak}
\def\be{\begin{equation}}
\def\ee{\end{equation}}
\def\bea{\begin{eqnarray}}
\def\eea{\end{eqnarray}}
\newcommand{\rf}[1]{(\ref{#1})}
\newcommand{\eq}[1]{Eq.~(\ref{#1})}

\def\appendix#1{
\addtocounter{section}{1}
\setcounter{equation}{0}
\renewcommand{\thesection}{\Alph{section}}
\section*{Appendix \thesection\protect\indent #1}
\addcontentsline{toc}{section}{Appendix \thesection\ \ \ #1}
}
\newcommand{\ie}{{\it i.e.}\ }

\newcommand{\half}{{\textstyle{1\over 2}}}
\renewcommand{\d}{{{\partial}}}
\newcommand{\p}{{\prime}}
\newcommand{\ra}{\rightarrow}
\def\la{\mathrel{\mathpalette\fun <}}
\def\ga{\mathrel{\mathpalette\fun >}}
\def\fun#1#2{\lower3.6pt\vbox{\baselineskip0pt\lineskip.9pt
\ialign{$\mathsurround=0pt#1\hfil##\hfil$\crcr#2\crcr\sim\crcr}}}

\title{
\hspace{4.2truein} {\small UBCTP 92-30}\\
 \hspace{4.2truein} {\small ITEP-YM-8-92}\\
\vspace{0.3truein}
{\bf Four-Fermion Theory and the Conformal Bootstrap }}
\author{Wei Chen$^\dagger$, Yuri Makeenko$^\ddagger$ and
 Gordon W.\ Semenoff$^\dagger$\\
\\
$^\dagger$ Department of Physics, University of British Columbia\\
Vancouver, British Columbia, Canada V6T 1Z1\\
\\
$^\ddagger$ Institute of Theoretical and Experimental Physics \\
117259 Moscow, Russia}

\vspace{0.2truein}
\date{October 1992}

\begin{document}
\maketitle

\vspace{0.1truein}
\begin{abstract}
We employ the conformal bootstrap to re-examine the problem of finding
the critical behavior of four-Fermion theory at its strong coupling
fixed point.  Existence of a solution of the bootstrap equations
indicates self-consistency of the assumption that, in space-time
dimensions less than four, the renormalization group flow of the
coupling constant of a four-Fermion interaction has a nontrivial fixed
point which is generally out of the perturbative regime. We exploit
the hypothesis of conformal invariance at this fixed point to reduce
the set of the Schwinger-Dyson bootstrap equations for four-Fermion
theory to three equations which determine the scale dimension of the
Fermion field $\psi$, the scale dimension of the composite field
$\bar{\psi}\psi$ and the critical value of the Yukawa coupling
constant. We solve the equations assuming this critical value to be
small. We show that this solution recovers the fixed point for the
four-fermion interaction with $N$-component fermions in the limit of
large $N$ at (Euclidean) dimensions $d$ between two and four. We
perform a detailed analysis of the $1/N$-expansion in $d=3$ and
demonstrate full agreement with the conformal bootstrap.  We argue
that this is a useful starting point for more sophisticated
computations of the critical indices.
\end{abstract}

\eop

\baselineskip=20.0truept

\newsection{Introduction}

Conformal symmetry governs the critical behavior of a quantum
field theory near the fixed points of its renormalization group flow
\cite{Pol,Mig71,Par72}.  It is most powerful in two spacetime
dimensions where the requirement of conformal invariance of
correlation functions leads to a solution of a quantum field theory in
terms of a few parameters \cite{bpz}.  In this Paper we shall show
that conformal symmetry is also useful in higher dimensions. In any
spacetime dimensions, it determines the form of the two and
three-point Green functions up to a few constants, the scaling
dimension of the field operators and the value of the coupling
constant at the fixed point.  When this information is combined with
Schwinger-Dyson equations, it can provide a powerful tool for the
analysis of field theories.  We shall apply the conformal bootstrap
technique to four-Fermion interactions in dimensions $2<d<4$.  An
application of this technique to $\lambda\phi^4$ theory
\cite{Mak79} and some preliminary work for spinors with four-Fermion
interactions \cite{Mak78} between 2 and 4 dimensions have been given
previously by Yu.  Makeenko.

Fermionic field theories with four-fermion interactions have a long
history.  A classic example is the Nambu-Jona-Lasinio model which was
proposed some thirty years ago by Nambu and Jona-Lasinio \cite{NJ}
(see also \cite{VL61}) as a quantum field theory which exhibits
dynamical chiral symmetry breaking for sufficiently strong attractive
fermion-anti-fermion interactions in four (space-time) dimensions.
The study of related models has recently seen a revival with
speculations that the symmetry breaking mechanisms which it describes
could be used in the standard model to describe the Higgs particle as
a top quark condensate \cite{top,bard}.

In the conventional perturbative expansion in powers of the coupling
constant, four-fermion interactions are not renormalizable (or are
irrelevant operators) in space-time dimensions greater than two.  This
can be seen by simple power-counting.  Consider the Euclidean action
in d dimensions
\be
S = \int d^dx \left(\bar\psi \hat\partial \psi -
\frac{G}{2}\left(\bar\psi\psi\right)^2\right)\;,
\label{act1}
\ee
where $\hat\partial\equiv\gamma\cdot\partial$ and $G$ is the (bare)
coupling constant.  The coefficient of the four-fermion coupling must
have engineering dimension $2 - d$:
\be
G={g^2/\Lambda^{d-2 }}\,.
\label{Gg}
\ee
Here, we have defined the dimensionless coupling constant $g^2$ and
have included powers of the ultraviolet cutoff $\Lambda$ to give the
vertex its correct dimension.  In $d = 2$ the coupling constant $G$ is
dimensionless and the model is renormalizable in the conventional
sense.  It was shown by Gross and Neveu \cite{GN} that it is
asymptotically free, has nontrivial scaling and contains
fermion-anti-fermion bound states.

In $d>2$, the four-fermion interaction is irrelevant in weak coupling
perturbation theory.  Because of the powers of cutoff
$1/\Lambda^{d-2}$ in the bare coupling, perturbative contributions to
the full four-fermion vertex vanish as the cutoff is sent to infinity.
For example, the contribution of each of the bubble diagrams in
Fig.~$1$ (the corresponding Feynman integral has ultraviolet
divergence $\sim\Lambda^{d-2}$) goes to zero like $1/\Lambda^{d-2}$ if
the external momentum is kept fixed at a finite value and
\mbox{$\Lambda\rightarrow\infty$}.  Thus, the four-fermion operator is
irrelevant at weak coupling: its contributions to four-fermion
scattering vanish and its contribution to relevant vertices can be
absorbed by local renormalizable counterterms.

On the other hand, it was conjectured long ago~ \cite{FP63} that the
four-fermion interaction could perhaps be relevant if, instead of a
simple perturbative expansion, one considered some nontrivial
resummation of infinite numbers of diagrams.  For example, if we
consider bubble graphs of the kind depicted in Fig.~$1$, the term with
n bubbles is
\be
n\hbox{-bubble graph} \sim
g^{2(n+1)}\Lambda^{(n+1)(2-d)}\left(\Gamma(\Lambda, p)\right)^n\;,
\ee
where $\Gamma(\Lambda, p)\sim\Lambda^{d-2}+\dots$ is the contribution
from a single bubble. The geometric sum over all bubble graphs of the
kind depicted in Fig.~$1$ has the form
\be
{\rm sum}~\sim~ {g^2\over
\Lambda^{d-2}-g^2\Gamma(\Lambda,p)}+{\rm crossed}\;.
\label{sum}
\ee
The interaction can be non-zero if
\be
g^2~=~ {\Lambda^{d-2}\over\Gamma(\Lambda,0)} +{\cal O}\left(\left(\rm
finite~ mass~ scales/\Lambda\right)^{d-2}\right)\,.
\ee
Thus, in order to produce non-zero interactions, the coupling constant
must be precisely tuned to this strong coupling (ultraviolet stable)
fixed point, within errors proportional to $\left(\rm finite~ mass
{}~scales/\Lambda\right)^{d-2}$.  At the critical point, (\ref{sum})
behaves like ${\rm sum}\sim 1/p^{d-2}$ so that the effective scale
dimension of the composite operator $\bar\psi\psi$ is 1 (in momentum
units) for any $d$.  When $d\neq2$ this differs from its classical
dimension, $(d-1)$.  Moreover, the dimension of the four-fermion
vertex is approximately two, rather than its classical dimension
$2(d-1)$.  It is this large anomalous dimension of composite operators
which makes the four-fermion interaction relevant.

The latter idea appeared in the work of Wilson \cite{W} who observed
that the sum of bubble diagrams in (\ref{sum}) is the leading order in
the large N expansion where N is the number of fermion species.  He
argued that, in dimensions $2<d<4$, as well as the infrared stable
fixed point at zero coupling, the four-fermion coupling has an
ultraviolet stable fixed point at a non-zero value of the coupling
constant $g^2_*>0$.  As $d\rightarrow2$ this fixed point moves to the
origin, $g^2_*\rightarrow 0$ and gives the asymptotic freedom of the
Gross-Neveu model \cite{GN}.  He also observed that at $d=4$ the
four-fermion theory is trivial and is described by $N$
non-interacting spinor fields and a free scalar field.

This led to the conjecture that, in $2<d<4$ the systematic large N
expansion of four-fermion theory is renormalizable in the conventional
sense that, in order to remove the cutoff dependence of physical
quantities as they are computed order by order in 1/N, it is necessary
to add local counterterms which have the same form as these already
occurring in the action~\cite{PG}\cite{RWP}\cite{HKK}.  That this is
indeed the case in $d=3$ has only recently been proven using the
renormalization group technique of constructive field
theory~\cite{CFVS}.

There have also been numerous recent efforts, both analytical
\cite{ZJ} and numerical \cite{HKK1}, to understand the critical
behavior of the four-fermion model around its fixed points.  The
numerical data in d=3 is in good agreement with analytical
calculations in the large $N$ expansion.  The ultraviolet fixed point
has the interesting feature that the critical indices there are much
different from those at the Gaussian infrared fixed point.

In order to implement the large N expansion it is convenient to
introduce an auxiliary scalar field, so that the action takes the form
\be
S = \int d^dx \left(\bar\psi_i \hat\partial \psi_i + \frac{1}{\sqrt
N}\phi\bar\psi_i\psi_i + \frac{1}{2G}\phi^2\right)\;,
\label{action}
\ee
where $i=1,\dots,N$.  The four-fermion interaction in \rf{act1} is
recovered by solving the equation of motion for $\phi$ or,
equivalently, performing the gaussian functional integral over $\phi$
in the partition function
\be
Z=\int d\psi d\bar\psi d\phi \exp( -S)\;.
\nonumber
\ee
The introduction of $\phi$ results in a Yukawa interaction which, as
we shall see, is essential to the method we shall introduce in this
paper. In the past this trick has been used to discuss~\cite{LM64} the
equivalence of the four-Fermi theory and a fermion-scalar field theory
coupled through Yukawa interactions when the wavefunction
renormalization constant for the $\phi$ field vanishes.  This is known
as the compositeness condition and, as has been suggested in some
recent literature \cite{bard}, can be used to reduce the number of
parameters in the fermion-scalar field theory, thereby giving the
model more predictive power.  Indeed, one could consider a
generalization of
\rf{action} which contains a kinetic term for the scalar field
$\frac{1}{2}\phi\Box\phi$ and possibly higher polynomials -- in fact
all of the relevant operators of the scalar fields theory.  Then if
one considers the renormalized kinetic term, $\frac{1}{2}{\cal
Z_\phi}\phi_R\Box \phi_R$ the compositeness condition can be achieved
if as a function of $g^2$ and other parameters, the equation ${\cal
Z\phi}(g^2,\dots)=0$ has a solution.  If, at this solution,
correlators of the field theory remain finite, the interacting fermion
theory which would result from eliminating the scalar fields is also
finite. (If the scalar field action has terms of higher than quadratic
order, the resulting fermion theory is rather complicated, possibly
non-polynomial.)

The action \rf{action} is invariant under the discrete chiral
transformation
\be
\psi_i ~\rightarrow~ \gamma_5\psi_i\;, ~~~
\bar\psi_i ~\rightarrow~ -\bar\psi_i\gamma_5\;,~~~
\phi ~\rightarrow -\phi\;.
\ee
where we consider the appropriate definition of $\gamma^5$ in less
than 4 dimensions.  (We use spinors with $2^{[d/2]}$ components, where
[d/2] is the smallest integer greater than or equal to d/2.  In any
integer dimensions such spinors have an analog of $\gamma^5$ which is
the product of the gamma matrices, $\prod_{\mu}\gamma^\mu$.)  This
chiral symmetry prevents the fermion from acquiring a mass.  When the
four-fermion interaction is attractive and strong enough, there is a
phase transition and the fermion acquires mass by spontaneous chiral
symmetry breaking. The corresponding critical coupling constant is
just the ultraviolet fixed point.  A natural order parameter for the
spontaneous chiral symmetry breaking is the expectation value of the
auxiliary field $\langle\phi\rangle$ which, in our convention, has the
dimension of mass.  The critical point where the phase transition
occurs can be approached from either the symmetric or the broken
phase.  Assuming the phase transition here is of second order, as
strongly suggested by numerical Monte-Carlo simulation \cite{HKK1},
the critical indices calculated in one phase must be equivalent to
those calculated in the other.

In the present paper, we shall use the conformal bootstrap to examine
four-fermion interactions in $2 < d < 4$ at the critical point.  The
conformal bootstrap was invented by Polyakov \cite{Pol}, Migdal
\cite{Mig71}, Parisi \cite{Par72} and many others in the early
seventies.

The basic idea is as follows.  As the transition point is approached,
various quantities either vanish (such as masses) or diverge and
long-range correlations appear.  Moreover, just at the critical point,
the correlation function of fluctuations takes the form
\be
\langle\delta s(x) \delta s(0)\rangle
\stackrel{|x|\rightarrow \infty}{\sim} |x|^{-(d-2-\eta)}\;.
\label{cr}
\ee
This is exactly the form of two-point correlation functions which is
implied by scale invariance, where the effective scale dimension of
the operator $\delta s(x)$ is $\frac{d-2-\eta}{2}$.

In fact, as was first observed in \cite{MS69}, in addition to scale
invariance, at the phase transition point the system is invariant
under the full group of special conformal transformations.  To see
this, consider the conformal current $K^\alpha_\mu(g^2,x)$ where $g^2$
is the effective coupling constant.  Then the divergence of the
conformal current is related to the beta function of $g^2$ through the
trace of the energy-momentum tensor:
\be
\partial_\mu K^\alpha_\mu (g^2,x)
\propto x^\alpha\theta_{\mu\mu}(g^2,x)\;,
\ee
which in turn is the divergence of the dilatation current
$D_\mu(g^2,x)$ and, therefore, proportional to the beta function:
\be
\theta_{\mu\mu}(g^2,x) =
\partial_\mu D_\mu (g^2,x)\propto \beta (g^2)\;.
\label{betafunction}
\ee
As usual, the last equation is understood in the sense of matrix
elements.  If more than one coupling constant is involved, the right
hand side of \eq{betafunction} would be a linear combination of all
the corresponding beta functions.

As well as the two-point function, which is actually fixed (up to the
critical exponent) by scale invariance, conformal symmetry also fixes
the form of three-point functions up to multiplicative constants.  The
multiplicative constants are the fixed point values of the renormalized
coupling constants --- by definition --- the zero point of their beta
function.

To determine the unknown exponents and multiplicative constant (fixed
point), one invokes the Schwinger-Dyson equations in bootstrap form --
the homogeneous Schwinger-Dyson equations.  The inhomogeneous term
arising from bare couplings in the Schwinger-Dyson equation for the
three-point vertex vanishes since it is multiplied by the inverse of
an infinite renormalization constant.  Since the bare vertex does not
have the correct form to satisfy the conformal ans\"atz, this
cancellation is in fact necessary for consistency of the assumption of
conformal symmetry. This is discussed in detail in \cite{Mig71}.  The
bootstrap equations for the three-vertex are closed and have a
conformal solution. They are free of ultraviolet and infrared
divergences, as a result of renormalization and are sufficient to
determine the scale dimension of the fields and the value of the
coupling constant at the fixed point.  In the present model, these are
the critical coupling constant and the scaling dimensions of the
fields $\psi$ and $\phi$.  If the hyperscaling hypothesis \cite{K} is
correct, all other critical exponents can be determined.

This paper is organized as follows. In Section~2 we analyze the $1/N$
expansion of four-fermion theory at $d=3$ to order $1/N$, demonstrate
renormalizability and review the calculation of the anomalous
dimensions of $\psi$ and $\phi$ to that order.  In Section~3 we assume
conformal invariance of the four-fermion theory at fixed point to
derive the set of bootstrap equations which determines the critical
indices. In Section~4 we consider the three-vertex approximation which
is legitimate when the index of the Yukawa vertex is small, calculate
the critical indices and show that results agree with the $1/N$
expansion.  In the Appendix we present a derivation of the Parisi
procedure for the case of the small index of Yukawa vertex.

\newsection{$1/N$ expansion}

In this section, we conduct a perturbative expansion of the theory in
$1/N$ with emphasis on the conformal structure.  For simplicity, we
restrict our computations to $d=3$, the generalization to any
dimensionality in $2< d< 4$ is straightforward.

\subsection{Bubble graphs}

Our convention is that each fermion has two components and the Dirac
matrices are Hermitean, namely, $\gamma_\mu\gamma_\nu + \gamma_\nu
\gamma_\mu = 2\delta_{\mu\nu}{\bf 1}$ and $\hbox{Tr}\,{\bf 1} =
2^{[d/2]} = 2.$ In the symmetric phase, fermions have no dynamical
mass.  The Feynman rules in the momentum space use the fermion
propagator and three-point vertex
\be
S_0(p) = \frac{1}{i\hat{p}}~~~\hbox{ and }~~~ \Gamma_0= -
\frac{1}{\sqrt N}\;,
\label{f1}
\ee
respectively.

The scalar, $\phi$, is not a dynamical but an auxiliary field.
However, there formally exists a bare $\phi$ `propagator', $G$,
corresponding to the last term in the action (\ref{action}).  $G$, the
only free parameter of the theory, has dimension of mass$^{-1}$.

On the other hand, at a fixed point all dimensionful parameters must
cancel from expressions for correlation functions.  Here, the
cancellation of the dimensional coupling constant $G$ occurs through
the the regularization and renormalization procedure.  If the theory
is regularized by introducing a cutoff $\Lambda$ which has the
dimension of mass, one can choose $1/G$ so that it exactly cancels
the terms in the scalar field self-energy that are linearly divergent,
$\sim\Lambda$.  Or, if the theory is regularized by a method such as
dimensional regularization where linear divergences are absent,
$1/G$ set to zero.  We can regard both of these as fine tuning of
the bare coupling constant $1/G$ to the ultraviolet stable fixed
point.

In this Section, we shall introduce a naive cutoff $\Lambda$ as a
regulator. In this case, we can define a {\it dimensionless} coupling
constant $g$ according to \eq{Gg}:
\be G = \frac{g^2}{\Lambda}\;.  \label{lam} \ee (In this Section, d=3.)

The bare $\phi$-propagator is a momentum independent and is
proportional to the coupling constant, $G$.  In the $1/N$ expansion,
the fermion bubble chains (Fig.~$2$) are of the same $0$-th order as
the bare $\phi$ propagator.  Therefore, the correct $\phi$ propagator
in the Feynman rules should be an effective one that sums over all
one-fermion-loop chains.  It is referred to as the dressed $\phi$
propagator and its inverse takes the form
\be
\Delta^{-1}_0 (p,\Lambda) = \frac{1}{G}
- \int^\Lambda\frac{d^3k}{(2\pi)^3}
\hbox{Tr} \frac{\hat{k}(\hat{k}+\hat{p})}{k^2(k+p)^2}
= \frac{\Lambda}{g^2} -\frac{\Lambda}{\pi^2} + \frac{1}{8}|p|\;.
\label{F1}
\ee

Setting $g=g_c$ so that to cancel the linear in $\Lambda$ term in
\eq{F1},  the h$1/G-\Lambda/\pi^2=0$,
we have the critical coupling constant at this order
\be
g_c = {\pi}\;,
\label{lamc0}
\ee
and the dressed $\phi$ propagator

\be
\Delta_0(p)=\frac{8}{|p|}\;.
\label{f2}
\ee

Now we see that, at the order $1/N^0$, the two- and three-vertices
given by Eqs.\rf{f1} and \rf{f2} have indeed a conformal structure, as
if the fields $\psi$ and $\phi$ both have dimension $1$ and the Yukawa
interaction is dimensionless.  While this value of the dimension of
the fermion field $\psi$ coincides with the canonical one at $d=3$, in
zeroth order in 1/N the scale dimension pg $\phi$ is 1, different from
the value 3/2 which is implied by the bare propagator.\footnote{Since,
for large momentum $\vert p\vert$, the momentum-space propagator of a
field with the scale dimension $l$ is $\sim [p^2]^{l-d/2}$ in a
$d$-dimensional space, the canonical {\it scale\/} dimension of the
auxiliary field is $d/2$ while that of a conventional dynamical scalar
field is $d/2-1$.}.  This is related to the shift of the scale
dimension of the composite operator $\bar\psi\psi$ to 1 after summing
the bubble graphs and tuning the bare coupling to the critical value
\rf{lamc0} which was discussed in Section 1.  Notice that the scale
dimension 1 coincides with the standard {\it mass\/} dimension of the
field $\phi$, so that there is no need of a mass parameter in
\eq{f2}.

\subsection{Primary divergences
at ${\cal O}(1/N)$ }

At the order ${\cal O}(1/N)$, we need to consider the diagrams in
Figs.~$3$ -- $6$. We shall see that the primary divergences are these
in the fermion self-energy and in the Yukawa vertex.  A
straightforward calculation of Fig.~$3a$ gives the regulated fermion
two-point vertex
\begin{eqnarray}
S^{-1}(p,\Lambda) &=& i\hat p - \frac{-8i}{N}
\int^\Lambda\frac{d^3 k}{(2\pi)^3}
\frac{\hat{k}+\hat{p}}{|k|(k+p)^2}\nonumber\\*
&=& i\hat{p} {\bf [} 1+\frac{2}{3\pi^2N}\ln(\frac{\Lambda^2}{p^2}) +
\hbox{finite}{\bf ]}\;;
\end{eqnarray}
and of Fig.~$4a$ the regulated three-point vertex
\begin{eqnarray}
-\Gamma(p_1,p_2, \Lambda) &=& - \frac{1}{\sqrt{N}} + \frac{8}{\sqrt{N^3}}
\int^\Lambda\frac{d^3 k}{(2\pi)^3}
\frac{(\hat{k}+\hat{p}_1)(\hat{k}+\hat{p}_2)}{|k|(k+p_1)^2(k+p_2)^2}
\nonumber\\*&=& - \frac{1}{\sqrt{N}}
{\bf [}1-\frac{2}{\pi^2N}\ln(\frac{\Lambda^2}{p^2_{\rm max}})
+\hbox{finite} {\bf ]}\;,
\label{gamma0}
\end{eqnarray}
where $p^2_{\rm max}$ is the largest of $p_1^2$ and $p_2^2$, with
$p_1$ and $p_2$ the incoming and outgoing fermion momenta,
respectively.

The diagrams in Fig.~$5$, which have a subdiagram with a fermion loop
and three boson legs.  The subdiagravanishes because of parity
symmetry.  We shall see in the next section that this form of the
vertex function is in fact prescribed by conformal invariance.

To render the above two- and three-point vertices finite, we need to
introduce local counterterms, Figs.~$3a^\p$ and $4a^\p$,
\begin{eqnarray}
\hbox{Fig.}3a^\p&=&- \frac{2}{3\pi^2N}
\ln(\frac{\Lambda^2}{\mu^2})\bar\psi\psi\;,
\label{count1}\\
\hbox{Fig.}4a^\p&=&- \frac{2}{\pi^2\sqrt{N^3}}
\ln(\frac{\Lambda^2}{\mu^2})\phi\bar\psi\psi\;,
\label{count2}
\end{eqnarray}
respectively, where a reference mass scale $\mu$ has been
introduced.  The corresponding renormalization constants, using
minimal subtraction, are
\begin{eqnarray}
Z_\psi(\Lambda) &=& 1-\frac{2}{3\pi^2N}\ln(\frac{\Lambda^2}{\mu^2})\;,
\label{zpsi} \\
Z_1(\Lambda)&=&1 + \frac{2}{\pi^2N}\ln(\frac{\Lambda^2}{\mu^2})\;.
\label{z1}
\end{eqnarray}
The anomalous dimension of the fermion field is
\be
\gamma_\psi = \frac{2}{3\pi^2N}\;.
\label{gpsi}
\ee
{}From \rf{gamma0}, we see that the three-point vertex, while
dimensionless at the tree level, develops an anomalous dimension,
$-2\gamma$, beyond the tree level.  We shall call $\gamma$ the index
of the Yukawa vertex.  At the order $1/N$,
\be
\gamma = -\frac{2}{\pi^2N}\;.
\label{g}
\ee

\subsection{A demonstration of renormalizability}

The renormalization of the operator $\phi^2$ is crucial.  By power
counting, the inverse of the two-point function
$\langle\phi\phi\rangle$ has engineering dimension one, instead of
two, which would be the case for a dynamical scalar field.  The self
energy for the scalar is linearly divergent, with the divergent terms
ocurring in Feynman diagrams cancelled order by order in the 1/N
expansion by adjusting the value of the critical coupling constant.
The remainder of the scalar self-energy can (and generally does)
contain individual integrals with non-local divergent terms like
$|p|\ln(\Lambda^2/p^2)$.  It should be rendered finite by the
insertion of the fermion wave-function renormalization and vertex
renormalization counterterms into the internal lines of lower order
diagrams.  Otherwise, non-local counterterms would be necessary to
cancel ultraviolet divergences and the 1/N expansion of the theory
would be non-renormalizable in the conventional sense.  In this
Section, we shall demonstrate that, to order 1/N, these nonlocal
divergences are indeed cancelled by local counterterms.

To see this, we consider the diagrams at the order ${\cal O}(1/N)$,
Fig.~$6$.  Figs.~$6a$ and $6b$ each has a subdiagram, the one-loop
fermion self-energy, that is logarithmically divergent, as calculated
above. Therefore accompanying Figs.~$6a$ and $6b$ are Fig.~$6a^\p$ and
$6b^\p$ containing the counterterm \rf{count1}.  Similarly, Fig.~$6c$
has a subdiagram, the one-loop vertex correction, that is also
logarithmically divergent, and therefore we need to calculate
Figs.~$6c^\p$ and $6c^\p$$^\p$, together with Fig.~$6c$.  Once these
counterterm diagrams have been taken into account all logarithmically
divergent terms cancel out between Figs.~$6a$ and $6a^\p$; Figs.~$6b$
and $6b^\p$; and among Figs.~$6c$, $6c^\p$ and $6c^\p$$^\p$.
Consequently, no momentum dependent counterterms are necessary.

The linearly divergent terms in the $\phi$ self-energy can be
cancelled by adjusting the critical coupling constant $g_c^2$, which
has been determined to leading order in the previous subsection. By
this analysis, we have confirmed the renormalizability of \rf{action}
at the order $1/N$.  In the language of multiplicative
renormalization, introducing counterterms implies introducing
renormalization constants for the vertices appearing in the action. In
the the present case, $Z_\phi$ for $\phi$ is not independent but takes
the form \cite{RWP}
\be
Z_\phi = \frac{Z_1^2}{Z_\psi^2}
= {\bf [}1+\frac{16}{3\pi^2N}\ln(\frac{\Lambda^2}{\mu^2}){\bf ]}\;.
\label{zphi}
\ee

The following explicit calculation confirms \rf{zphi}:
\begin{eqnarray}  & &  \hbox{Fig.}5a+\hbox{Fig.}5b+\hbox{Fig.}5c
\nonumber \\*
&= & -\frac{8}{N} \int\frac{d^3 k}{(2\pi)^3}\frac{d^3
q}{(2\pi)^3}\hbox{Tr} {\bf
[}2\frac{\hat{k}(\hat{k}+\hat{p})\hat{k}(\hat{k}+\hat{q})}
            {k^4(k+p)^2(k+q)^2q} +
\frac{\hat{k}(\hat{k}+\hat{p})(\hat{k}+\hat{p}+\hat{q})(\hat{k}+\hat{q})}
            {k^2(k+p)^2(k+q)^2(k+p+q)^2q} {\bf ]} \nonumber\\*
&=& -\frac{2}{N}\int^\Lambda\frac{d^3 q}{(2\pi)^3}
{\bf [} \frac{2}{|q+p|q}
        +\frac{q\cdot p}{q^2|q+p|p}
        +\frac{p}{q\cdot p|q+p|} {\bf ]}\nonumber\\*
&=&\frac{1}{\pi^2N}{\bf [} - 2\Lambda
+ \frac{2}{3}p\ln(\frac{\Lambda^2}{p^2}) +\hbox{const}\times p {\bf ]}\;.
\end{eqnarray}
By using (\ref{F1}), we have
\begin{eqnarray}
\Delta^{-1}(p, \Lambda) = \frac{\Lambda}{g^2}
-{\bf [} \frac{\Lambda}{\pi^2} - \frac{p}{8}{\bf ]}
-\frac{1}{\pi^2N}{\bf [} - 2\Lambda
+ \frac{2}{3}p\ln(\frac{\Lambda^2}{p^2}) +\hbox{const}\times p {\bf ]}\;.
\end{eqnarray}
Choosing $g=g_c$ so that
$\Lambda/g_c^2 - (1-2/N)\Lambda/\pi^2 = 0$, we have the
critical coupling constant at the order $1/N$
\be
g_c=\pi(1+\frac 1N)\;.
 g^2_c = \frac{1}{\pi^2}(1-\frac{2}{N}).
\label{lamc1}
\ee
 The renormalized scalar two-point function is
\be
 \Delta^{-1}(p)
= Z_\phi(\Lambda)\Delta^{-1}(p, \Lambda)
= \frac{p}{8}{\bf [}1-\frac{16}{3\pi^2N}\ln(\frac{\mu^2}{p^2}){\bf ]}\;,
\ee
where we have used the $\phi$ wavefunction renormalization constant
\rf{zphi}. The anomalous dimension of $\phi$ is
\be
\gamma_\phi = -\frac{16}{3\pi^2N}\;.
\label{gphi}
\ee

{}From \rf{gpsi}, \rf{g} and \rf{gphi}, it can be checked that
the index of Yukawa interaction and the anomalous dimensions of
$\psi$ and $\phi$ at the order ${\cal O}(1/N)$ satisfy a relation

\be
\gamma = \gamma_\psi + \frac{1}{2}\gamma_\phi\;.
\label{vertexindex}
\ee
As we shall show in the next section, this is not an accident but a
result of conformal symmetry. The point is that the renormalization
constants in the conformal theory are equal to their fixed point
values:
\begin{eqnarray}
Z_\psi(\Lambda)&=&\Big(\frac{\Lambda^2}{\mu^2}\Big)^{\gamma_\psi}\;,
\label{zpsiconf} \\
Z_1(\Lambda)&=&\Big(\frac{\Lambda^2}{\mu^2}\Big)^{\gamma}\;,
\label{z1conf} \\
Z_\phi(\Lambda)&=&\Big(\frac{\Lambda^2}{\mu^2}\Big)^{\gamma_\phi}\;.
\label{zphiconf}
\end{eqnarray}
The expressions \rf{zpsi}, \rf{z1} and \rf{zphi} which are explicitly
calculated at the order ${\cal O}(1/N)$ can be viewed therefore as the
${\cal O}(1/N)$ terms of the expansion of \rf{zpsiconf}, \rf{z1conf}
and \rf{zphiconf} in $1/N$ with $\gamma_\psi$, $\gamma$ and
$\gamma_\phi$ given by Eqs.~\rf{gpsi}, \rf{g} and \rf{gphi},
respectively.


\newsection{Conformal bootstrap}

In this section, we derive the bootstrap equations for the system with
four-fermion interactions in $d$ dimensions.  The result is three
equations which determine two anomalous dimensions and the value of
the critical coupling constant.  We start with a systematic analysis
of the conformal structure of a general theory involving fermions and
bosons with Yukawa interaction(s).

\subsection{Conformal structure}

Let $l$ and $b$ be the scaling dimensions of the primary
spinor and scalar fields, respectively.
Under a scaling transformation of the coordinates, $x \rightarrow \rho x$,
the transformation of the fields are defined as

\be
\psi ^\p(\rho x)=\rho^{-l}\psi (x)\;,~~
\bar{\psi}^\p (\rho x)=\rho^{-l}\bar{\psi} (x)\;,~~
\phi^\p (\rho x)=\rho^{-b}\phi (x)\;.
\ee

Moreover, the special conformal transformation
of the fields are defined as:
\begin{eqnarray}
\psi (x)~&\rightarrow&~\psi ^\p(x^\p)
=\sigma_x^{l-1/2}(1+\hat{t}\hat{x})\psi(x)\;,
\label{psi}\\
\bar{\psi}(x)~&\rightarrow&~\bar{\psi}^\p(x^\p)
=\sigma_x^{l-1/2}\bar{\psi}(x)(1+\hat{x}\hat{t})\;,
\label{psib}\\
\phi (x)~&\rightarrow&~\phi ^\p(x^\p)
=\sigma_x^b \phi (x)\;,
\label{phi}
\end{eqnarray}
where $\sigma_x \equiv \sigma_t(x)
=|\frac{\partial x}{\partial x^\p}|^{1/d}=
1+2t\cdot x+t^2x^2,$
under a special conformal transformation of coordinates,
parameterized with a constant vector $t_\mu$,
\be x_\mu~\rightarrow ~x^\p_\mu
= \frac{x_\mu + t_\mu x^2}{1+2t\cdot x +t^2x^2}\;.
\label{conf}
\ee
Scaling and special conformal transformations together with
translations and Lorentz rotations comprise the conformal group which
has $(d+1)(d+2)/2$ generators in $d$ dimensions.

The transformations of the primary fields determine the
transformations of the Green functions.  For instance, under the
conformal transformation \rf{conf}, the two- and three-point functions
transform as
\begin{eqnarray}
G^\p(x^\p-y^\p)=\langle T
\psi^\p(x^\p)\bar{\psi}^\p(y^\p)\rangle
=(\sigma_x\sigma_y)^{l-1/2}(1+\hat{t}\hat{x})G(x-y)
(1+\hat{y}\hat{t})\;,\label{conff}\\
D^\p(x^\p-y^\p)=\langle T\phi ^\p(x^\p)\phi
^\p(y^\p)\rangle
=(\sigma_x\sigma_y)^b D(x-y)\;,\label{confff}\\
G_3^\p(x^\p,y^\p;z^\p)=\langle T\phi ^\p(z^\p)\psi
^\p(x^\p)\bar{\psi}^\p(y^\p)\rangle
=\sigma_z^b
(\sigma_x\sigma_y)^{l-1/2} (1+\hat{t}\hat{x})G_3(x,y;z)(1+\hat{y}\hat{t})\;.
\label{conffff}
\end{eqnarray}
The unique (modulo normalization) solution to the special conformal
transformations $\rf{conff}$ -- $\rf{conffff}$ is
\begin{eqnarray}
G(x-y) &=&
\frac{1}{4^{h-l}\pi^hi\tilde{N}(l)} \frac{\hat{x}-\hat{y}}{[(x-y)^2]^{l
           +1/2}}\;,
\label{S}\\ D(x-y)
&=& \frac{1}{4^{h-b}\pi^hiN(b)}
           \frac{1}{[(x-y)^2]^b}\;,
\label{D}\\
G_3(x,y;z)&=&C \frac{\hat{x}-\hat{z}}{[(x-z)^2]^{b/2 +1/2}}
\frac{\hat{z}-\hat{y}}{[(z-y)^2]^{b/2 +1/2}}
\frac{1}{[(x-y)^2]^{l - b/2}}\;,
\label{G}
\end{eqnarray}
where $h=d/2$, and
\be
\tilde{N}(\tau) = \frac{\Gamma (h-\tau+1/2)}{\Gamma (\tau+1/2)},\;
N(\tau) = \frac{\Gamma (h-\tau)}{\Gamma (\tau)}.
\ee
Note that, unlike the case of d=2 where the conformal group has an
infinite number of generators, the finite dimensional conformal
symmetry in $d>2$ restricts but does not fix the functional forms of
higher point correlation functions.

\rf{S} and \rf{D} are normalized so that their Fourier
transformations take a simple form
\begin{eqnarray}
 G(p) &=& \frac{1}{i\hat{p}}(p^2)^{l-h+1/2}\;,
\label{gg}\\
 D(p) &=& \frac{1}{p^2} (p^2)^{b-h+1}\;.
\label{dd}
\end{eqnarray}
Amputating the external legs
\footnote{This amputation in conformal theories replaces the dimensions by
`shadow' dimensions $l^*=d-l$, $b^*=d-b$.}
from the three-point function \rf{G},
we obtain the conformal three-point vertex (depicted in Fig.~$7$)
\bea
\Gamma(x,y;z)&=& g_* \frac{\Gamma(h)}{4^\gamma\pi^d}
N(\gamma)\tilde N^2(b/2)N(l-b/2) \nonumber \\
& & \cdot \frac{\hat{x}-\hat{z}}{[(x-z)^2]^{h-b/2+1/2}}
\frac{\hat{z}-\hat{y}}{[(z-y)^2]^{h-b/2+1/2}}
\frac{1}{[(x-y)^2]^{h-l +b/2}}\;,
\label{Gama}
\eea
which is normalized so that the Fourier transformation of \rf{Gama} is
\be
\Gamma(p_1,p_2)=g_*\Gamma(h)N(\gamma)
\int \frac{d^dk}{\pi^h}\frac{\hat k +\hat p_1}{[(k+p_1)^2]^{b/2+1/2}}
\frac{\hat k +\hat p_2}{[(k+p_2)^2]^{b/2+1/2}}
\frac{1}{[k^2]^{l-b/2}}\;,
\label{gamap}
\ee
where $g_*$ is the unknown
dimensionless fixed-point coupling constant and
\be
\gamma = l + b/2 - h\;,
\label{index}
\ee
is defined as the index of the vertex. By dimensional analysis, it is
easy to check that the dimension of the vertex is $-2\gamma$.

In the previous section, we have introduced in $d=3$ the anomalous
dimensions $\gamma_\psi$ and $\gamma_\phi$ by Eqs.~\rf{zpsiconf} and
\rf{zphiconf}. They are related to the scale dimensions $l$ and $b$ as
follows:
\begin{eqnarray}
 l&=&h - \half + \gamma_\psi\;,
\label{lvsgamma} \\
 b&=&1+\gamma_\phi\;.
\label{bvsgamma}
\end{eqnarray}
When these formulas are substituted into \eq{index}, it recovers
\eq{vertexindex} which is obtained in the previous section by explicit
calculations at the order ${\cal O}(1/N)$. It is worth also noting that when
\rf{lvsgamma} and \rf{bvsgamma} with $\gamma_\psi=\gamma_\phi=0$ are
substituted into \rf{gamap} the result coincides in $d=3$ with the
vertex function \rf{gamma0} which explains its conformal origin.

\subsection{Bootstrap equations}

The critical coupling $g_*$ and the indices $b$ and $l$ are determined
by the set of homogeneous Schwinger-Dyson equations, which we shall
now derive. To simplify notations, we introduce operators $S\Gamma$
and $S^\p\Gamma$ as
\be
 \langle p_1 | S\Gamma|p_2\rangle = G(p_1) \Gamma(p_1,p_2) G(p_2)\;, ~~
 \langle p_1 | S^\p\Gamma|p_2\rangle = G(p_1) \Gamma(p_1,p_2) D(p_2-p_1)\;,
\ee
where the matrix multiplication over the spinor indices is implied.
Here, $S$ and $S'$ are the operation of attatching two fermion or a
fermion and boson leg to the vertex function, respectively.  Then the
set of Schwin\-ger-Dyson equations reads
\bea
\Gamma &=& g_0 + \Gamma S K\;,
\label{ds1}\\
\Sigma \equiv G_0^{-1}-G^{-1}&=& g_0 \,S^\p \,\Gamma\;,
\label{ds2}\\
\Pi \equiv D_0^{-1}-D^{-1}&=& - g_0 \,N\,\hbox{Tr}\,(S\Gamma)\;,
\label{ds3}
\eea
where  \ Tr \ is the trace over the spinor indices and $K$ stands for the
standard Bethe--Salpeter kernel.

The bare coupling $g_0$ in \eq{ds1} must vanish so that this equation
is consistent with the conformal ansatz. This is realized by
renormalization \cite{Mig71}.
Moreover, since the three vertex is
fixed by conformal invariance, both sides of \rf{ds1} must have the
same dependence on the space-time coordinates.  Therefore the vertex
bootstrap equation takes a simple form
\be
1=g^2_* f(l,l,b;g_*)\;,
\label{confve}
\ee
where $f$, named the vertex function, is a function solely of the
scaling dimensions carried by the legs and of the critical coupling.
The vertex function, to be explored to some extent in the next
section, plays a key role in the conformal bootstrap approach.  
calculated below.  
lines is shown 
$l$ or $b$).

However, one cannot immediately set the bare coupling $g_0$ to zero in
Eqs.~\rf{ds2} and \rf{ds3}.  The reason is that the quantities that
$g_0$ multiplies in \rf{ds2} and \rf{ds3} are divergent. Therefore the
right hand side of \rf{ds2} and \rf{ds3} has the form $0\cdot\infty$.

To deal with this uncertainty, we differentiate both sides of
Eqs.~\rf{ds2} and \rf{ds3} with respect to the external momentum k.
Denoting $\frac{\d \Sigma}{\d k_\alpha}\equiv
\Sigma_\alpha$ and $\frac{\d \Pi}{\d k_\alpha} \equiv \Pi_\alpha \,$
and replacing $g_0$ with $\Gamma - \Gamma S K$ and $\Gamma - \Gamma
S^\p K^\p$, we have
\bea
\Sigma_\alpha &=& ( \Gamma- \Gamma S^\p K^\p) S^\p_\alpha \Gamma+
( \Gamma- \Gamma S^\p K^\p) S^\p \Gamma_\alpha\; ,\label{newpropagator1}\\
\Pi_\alpha &=& - N\,\hbox{Tr}{\bf [}( \Gamma- \Gamma S K) S_\alpha \Gamma+
( \Gamma- \Gamma S K) S \Gamma_\alpha {\bf ]}\; ,
\label{newpropagator2}
\eea
where $K^\p$ denotes the boson-fermion Bethe-Salpeter kernel. The skeleton
expansions of $K$ and $K^\p$ are depicted in Fig.~$8$.
The vertex bootstrap equation now is
\be
\Gamma=\Gamma SK\equiv \Gamma S^\p K^\p\;.
\label{vertexbootstrap}
\ee

Now we see that the first term on the right hand side of
Eqs.\rf{newpropagator1} and \rf{newpropagator2} vanishes as the
bracket is zero, according to \rf{vertexbootstrap}, and $S^\p_\alpha
\Gamma$ and $S_\alpha \Gamma$ are finite.  However, the second term of
Eqs.~\rf{newpropagator1} and \rf{newpropagator2} is still the form
$0\cdot\infty$.

\subsection{Parisi approach}

To remove the remaining uncertainty in \rf{newpropagator1} and
\rf{newpropagator2}, we use a
regularization proposed by Parisi~\cite{Par72}.  The idea is to shift
slightly the dimension of {\it one} of the operators in the Green
functions:
\be
l\ra l^\p=l+\epsilon\;\;\;,\hbox{ or \ \ }b\ra b^\p=b+\epsilon\;.
\ee
The corresponding change of the expression on the right hand side of
the vertex bootstrap equation is
\bea
\Gamma^\epsilon S^\p K^\p &=& g^2_*f(l+\epsilon,l,b;g_*)
\Gamma^\epsilon \;, \\
\Gamma^\epsilon S K &=& g^2_*f(l,l,b+\epsilon;g_*)
\Gamma^\epsilon\;,
\eea
where $f$ is defined similarly to the right hand side of \eq{confve}.
Using
\eq{confve} we obtain
\bea
\Sigma_\alpha&=& -\left.g^2_* \frac{\d f(l^\p,l,b;g_*)}{\d
l^\p}\right|_{l^\p=l} \epsilon \,\Gamma^\epsilon S^\p \,\Gamma_\alpha\;,
\label{finally1}\\
\Pi_\alpha&=& -N\left. g^2_* \frac{\d f(l,l,b^\p;g_*)}{\d
b^\p}\right|_{b^\p=b} \epsilon \,
\hbox{Tr}\,{\bf [}\Gamma^\epsilon S\, \Gamma_\alpha{\bf ]}\;.
\label{finally2}
\eea
Notice that, in this procedure, the dimension of only one {\it
external\/} line is analytically continued in the equation.

The calculations of $\Gamma^\epsilon S^\p \Gamma_\alpha$ and
$\Gamma^\epsilon S \Gamma_\alpha$ are performed similarly to those of
the scalar $\Phi^3$ theory~\cite{Par72} and of $\Phi^4$
theory~\cite{Mak79}. In the fermion model, a useful formula is
\cite{Sym72}
\begin{eqnarray}
\int \frac{dz}{\pi^h} \frac{(\hat{x}_1-\hat{z})(\hat{z}-\hat{x}_2)}
{[(x_1-z)^2]^{\delta_1+\half}[(x_2-z)^2]^{\delta_2+\half}
[(x_3-z)^2]^{\delta_3}} \nonumber \\ =
\frac{\hat{x}_{13}}{[x^2_{13}]^{h-\delta_2+\half}}
\frac{\hat{x}_{23}}{[x^2_{23}]^{h-\delta_1+\half} }
\frac{\tilde{N}(\delta_1)\tilde{N}(\delta_2)N(\delta_3)}
{[x^2_{12}]^{h-\delta_3}}\;.
\label{identity}
\end{eqnarray}
It is valid providing
\be
\delta_1+\delta_2+\delta_3=d\;.
\label{condition}
\ee
The meaning of these formulas is very simple:
if one performs the conformal transformation \rf{conf}
of $x$'s, the left hand side of \eq{identity} obeys \eq{conffff} providing
\eq{condition} for $\delta$'s is fulfilled. Therefore, the integral is
determined (modulo an $x$-independent factor) by the conformal invariance to
be the right hand side of \eq{identity}.

Using \eq{identity} in calculating $\Gamma^\epsilon S^\p\Gamma_\alpha$
 and $\Gamma^\epsilon S \Gamma_\alpha$ for each of the integrals (the all are
of the type \rf{identity})
and picking up the terms $\sim \epsilon^{-1}$, we
rewrite Eqs.~\rf{finally1} and \rf{finally2} as~\cite{Mak78}
\bea
1=- \frac{g^4_*}{(4\pi)^h} \Gamma(h) N(\gamma)\tilde{N}(l)N(d-l)
\tilde{N}^2(b/2)N(l-b/2)
\left.\frac{\partial f(l^{\prime},l,b;g_*)}{\partial l^\prime /2}
\right|_{l^\prime=l}\;,
\label{fParisi}\\
1=N\,\hbox{Tr}\,{\bf 1}\frac{g^4_*}{(4\pi)^h}
\Gamma(h)N(\gamma)N(b)N(d-b)N(l-b/2)\tilde{N}^2(b/2)
\left.\frac{\partial f(l,l,b^\prime;g_*)}{\partial b^\prime/2}
\right|_{b^\prime=b}\;.\nonumber\\
& &
\label{bParisi}
\eea
Eqs.~\rf{fParisi} and \rf{bParisi} together with
\eq{confve} determine the critical indices.
All the three equations depend on the vertex function $f(l,l,b)$,
which we shall discuss in the next section.
%
%

\newsection{Three-point vertex function}

The skeleton expansion of the Bethe-Salpeter kernel $K$ (or $K^\p$) in
the vertex bootstrap equation \rf{vertexbootstrap} contains an
infinite number of terms. If the critical coupling is reasonably
small, the three vertex approximation makes sense.  Then the vertex
equation has a solution with small $g_*$ for $\gamma\ll 1$.  The point
is that the integral in the vertex equation is slowly convergent if
$\gamma\ll 1$ and is therefore large ($\sim 1/\gamma$).  This enables
it to cancel the small factor of $g_*^2$ in the bootstrap equation.

\subsection{Momentum-space analysis}

It is convenient to perform calculation in the momentum space where the term
$\sim 1/\gamma$ comes from the region with large integration momentum.
The Fourier image of the
conformal vertex \rf{Gama} has been given in \rf{gamap}. The
integration on the right hand side of \rf{gamap} has an expression
\cite{FGGP74} of the Appel functions $F_4$ in general, and
takes a simple form for $\gamma \ll 1$. We shall derive the corresponding
formula directly analyzing the integral on the right hand side of \eq{gamap}.

Let us first note that the coefficient in front of the integral is
$N(\gamma) \approx \gamma$ for small $\gamma$, so that one has to peak up the
term $\sim 1/\gamma$ for the vertex to be of order $1$. It is easy to see that
this term comes from the region of integration with $|k| \ga \hbox{max}
\{|p_1|, |p_2|\}$.
(Recall that $|p_1-p_2| \la \hbox{max} \{|p_1|, |p_2|\}$ in
an Euclidean domain.) One gets
\bea
\int \frac{d^dk}{\pi^h}\frac{\hat k +\hat p_1}{[(k+p_1)^2]^{b/2+1/2}}
\frac{\hat k +\hat p_2}{[(k+p_2)^2]^{b/2+1/2}}
\frac{1}{[k^2]^{l-b/2}} \nonumber \\ = \int^\infty_{max\{p_1^2,p^2_2\}}
\frac{dk^2}{[k^2]^{1+\gamma}} =
\frac{1}{\gamma\,\left({\textstyle max}\{p_1^2,p_2^2\}\right)^\gamma}\;.
\label{gamapmax}
\eea
and
\be
\Gamma(p_1,p_2)=g_* \frac{1}{\left({\textstyle max}
\{p_1^2,p_2^2\}\right)^\gamma}\;.
\label{maxvertex}
\ee

This dependence of the three-point vertex solely on the largest momentum is
typical for logarithmic theories in the ultraviolet region where one can put,
say, $p_1=0$ without changing the integral with logarithmic accuracy. This is
valid, however, if the integral is fast convergent
in infrared regions. For our case this means that the dimensions $l$ and $b$
should be far away from the values at which the integral on the left hand
side of \eq{gamapmax} is infrared divergent. For instance if $l-b/2 \approx
h$, the term of the form
\be
\hbox{infrared term} = \frac{1}{l-b/2-h}
\frac{\hat p_1 \hat p_2}{[max \{p_1^2,p_2^2\}]^{b/2+1/2}
[min\{p_1^2,p_2^2\}]^{l+1/2-h}}\;,
\ee
were appear in the integral which would depend therefore both on the maximal
and on the minimal momenta.

 In particular,
 for $\gamma=-2/(N\pi^2)$ \eq{maxvertex} reproduces to order ${\cal O}(1/N)$
 Eq.(2.??) which was obtained by direct calculations. For this reason our
 analysis of the case of small $\gamma$ is very similar to that of Section~2.

Let us explicitly write down the dependence of the
vertex function $f$ on $g_*$:
\be
f(l,l,b;g_*)= g^2_*
f_3(l,l,b)
+ g^2_* f_5(l,l,b) + \ldots\,,
\label{vertexpansion}
\ee
where $f_3(l,l,b)$ is
associated with the three-point vertex contribution on the right hand side of
the
vertex bootstrap equation \rf{vertexbootstrap}, $f_5(l,l,b)$ with the
five-vertex terms
and so on. If one restricts oneself to the three-point vertex term, the
resulting
equation is depicted in Fig.~$9$. This truncation is referred as the
three-vertex approximation.

It is not difficult to calculate $f_3$ if $\gamma\ll1$. In order to
pick up the term
$\sim 1/\gamma$ one performs quite similar calculations to the logarithmic
case. Let $p_2^2 \ga p_1^2$. Then  the three
vertex equation  reads
\bea
\frac{1}{[p_2^2]^\gamma}&
=& - g^2_* \int \frac{d^dk}{(2\pi)^d}
\frac{1}{[{\textstyle max}\{p_1^2,k^2\}]^\gamma}
\frac{\hat{k}+\hat{p}_1}{[(k+p_1)^2]^{h-l+1/2}}\nonumber\\ & &\cdot
\frac{1}{[{\textstyle max}\{p_2^2,k^2\}]^\gamma}
\frac{\hat{k}+\hat{p}_2}{[(k+p_2)^2]^{h-l+1/2} }
\frac{1}{[{\textstyle max}\{p_2^2,k^2\}]^\gamma}
\frac{1}{[(k^2)]^{h-b}}
\label{threevertexeq}
\eea
where the momenta are depicted in Fig.~$9$.

The term $\sim 1/\gamma$ on the right hand side\ of \eq{threevertexeq}
comes from the domain where the integration momentum
$k^2 \ga p_2^2$.
Then we have finally
\be
f_3(l,l,b) = -\frac{1}{(4\pi)^h\Gamma(h)} \int_{p_2^2}^\infty
\frac{dk^2}{[k^2]^{1+\gamma}}
=-\frac{1}{(4\pi)^h\Gamma(h)\,\gamma}\,.
\label{f_3}
\ee
It has been verified simultaneously
that the momentum dependence (\ref{maxvertex})
is reproduced by the three vertex equation at small $\gamma$. This is
a consequence of conformal invariance which fixes as is mentioned above the
coordinate (or momentum) dependence of the three-point vertex.

To apply the procedure described in the previous section, we
repeat the calculation above with the dimension of one of the external
lines shifted and have
\bea
f_3(l^\p,l,b)
&=&-\frac{1}{(4\pi)^h\Gamma(h)\,\gamma^\p}\;,\;\;\;
\gamma^\p=\frac{l^\prime+l+b-d}{2}\,;\label{f_3p1}\\
f_3(l,l,b^\p)
&=&-\frac{1}{(4\pi)^h\Gamma(h)\,\gamma^\p{^\p}}\;,\;\;\;
\gamma^\p{^\p}=\frac{l+l+b^\p-d}{2}\,.
\label{f_3p2}
\eea
Such a modification is prescribed by the dimensional analysis.

Substituting Eqs.~\rf{f_3p1} and \rf{f_3p2} into Eqs.~\rf{fParisi}
and \rf{bParisi} for
propagator, we arrive at the algebraic equation set
 that determines the critical indices in
the three vertex approximation:
\bea
1&=&-\frac{g^2_*}{(4\pi)^h\Gamma(h)\,\gamma},
\label{ve}\\
1&=&\frac{g^2_*}{(4\pi)^h}\tilde{N}(b/2) N(l-b/2) \tilde{N}(d-l)\,,
\label{fe}\\
1&=&-N\,\hbox{Tr}\,{\bf 1} \frac{g^2_*}{(4\pi)^h}\tilde{N}^2(b/2)N(d-b)\,.
\label{be}
\eea

\subsection{The solution at large $N$}

We look for a solution to the algebraic
equations (\ref{ve}) -- (\ref{be})
for small $\gamma = l + b/2 - h$.
Let us assume the scaling dimension of $\psi$ is close
to the canonical value, $l \approx h-1/2$, \ie $\gamma_\psi$ which is defined
by \eq{lvsgamma} is small.
In order that
$\gamma = \gamma_\psi+(b-1)/2\ll 1$, the scaling dimension of $\phi$
should be $b\approx 1$, \ie $\gamma_\phi$ which is defined by \eq{bvsgamma}
should also be small.

{}From Eqs.~\rf{ve} and \rf{be} we have
 Let us first equal the r.h.s.'s of Eqs.~\rf{be} and \rf{fe} for $b\approx 1$
 and $l-h+1/2\ll1$. Solving the equation, we get

\be
l=h-1/2 + \frac{1}{N\,\hbox{Tr}\,{\bf 1}} \frac{\Gamma(2h-1)\sin{[\pi(h-1)]}}
{\Gamma(h-1)\Gamma(h+1)\pi }\,,
\label{Delta}
\ee

and from  Eqs.~\rf{ve} and \rf{fe}

\be
\gamma=-
\frac{1}{N\,\hbox{Tr}\,{\bf 1}} \frac{\Gamma(2h-1)\sin{[\pi(h-1)]}}
{\Gamma^2(h)\pi }\,,
\label{gamma}
\ee
which gives
\be
b=1-
\frac{2}{N\,\hbox{Tr}\,{\bf 1}} \frac{\Gamma(2h)\sin{[\pi(h-1)]}}
{\Gamma(h)\Gamma(h+1)\pi }\,.
\label{b-1}
\ee

Now we see that the scaling dimensions of $\psi$ and $\phi$ are such that
$\gamma_\psi$ and $\gamma_\phi $  are  $\sim 1/N$
and therefore are indeed small in the large $N$ limit.
Moreover, since $g^2_*\sim1/N$, it is verified that for large $N$
the three-vertex approximation is a good one.

Substituting $d=3$ and $\hbox{Tr}\,{\bf 1} = 2$ into  \rf{ve},
\rf{Delta} and \rf{b-1}, we  obtain
\bea
l &=&1+\frac{2}{3\pi^2N}\,, \label{d=3} \\ b
&=&1-\frac{16}{3\pi^2N}\,.
\label{D=3}
\eea
These values of scaling dimensions coincide
with those,  given by \rf{gpsi} and \rf{gphi}, which are calculated in
Section~2 by  the $1/N$ expansion while the (nonuniversal) value of
$g_*$, which is determined by \eq{ve},
is to be compared to \rf{lamc1}.

Notice that the value of $\gamma_\psi$ which is given by the last term
on the right hand side of \eq{Delta} is positive for $2<d<4$. In other
words, the dimension $l$ is bigger than the canonical value. This is
in agreement with the Lehman theorem which says that the momentum
space propagator in an interacting theory without ghosts should grow
with momentum faster than the free one. For the auxiliary field $\phi$
the Lehman theorem is not applicable since there is no pole term in
the spectral representation of the propagator, so that the dimension
$b$ can be arbitrary in four-fermion theory. However, if we were
consider instead a field theory with dynamical scalars coupled to
fermions through Yukawa interaction, it would be described by exactly
the same bootstrap equations with the solution given by
\rf{Delta} -- \rf{b-1}. In this case the Lehman theorem for the scalar field
would say that $b$ should be bigger than $h-1$, \ie than the canonical
value for a dynamical scalar field. We see that \eq{b-1} obeys this
restriction providing $d<4$. Therefore, our solution is applicable to
Yukawa theory as well as to four-fermion theory. This confirms that
the two theories are equivalent, as was discussed in Section~1, for
$2<d<4$.

\subsection{Estimating high-vertex contributions}

Let us stress once more that the system of equations \rf{ve} --
\rf{be} is valid provided no infrared divergences emerge in the
vertex bootstrap equation.  This is indeed the case for the values of
$l$, $\gamma$ and $b$ given by Eqs.~\rf{Delta} --
\rf{b-1} (or \rf{d=3} and \rf{D=3} for $d=3$).

The three vertex approximation works if five etc.\ vertex
contributions are small.  It is easy to estimate the order in $1/N$ of
these higher-vertex terms which have been disregarded.  This can be
done as follows.

The index of integrals in the Bethe-Salpeter kernel (which coincides
with that for a four-fermion vertex) is
\be
t=2l-h\;.
\ee
If $t\gg \gamma$, five etc.\ vertex contributions are suppressed as
\be
\frac{\lambda^2}{t}\sim \frac{\gamma}{t}\;.
\ee
This is true in particular for $d=4-\epsilon \,\,\,(\epsilon\ll1)$
when $\gamma\sim\epsilon$ but $t\sim1$.  However, if $t\sim\gamma$,
all parquet graphs are of the same order and should be summed up (like
four-boson interaction in $4-\epsilon \,\,\, (\epsilon\ll1)$. Another
example is Gross-Neveu model in $2+\epsilon\,\,\,(\epsilon\ll1)$ when
the $s$-channel graph is compensated by the $u$-channel one so that
the three vertex approximation works while $t\sim\gamma$.

\section{Conclusions}

The method of conformal bootstrap turned out to be very useful for
calculating the critical indices for the four-Fermi interaction at the
fixed point. From the viewpoint of the $1/N$ expansion this method
corresponds to a resumming of the perturbative series, so that at each
order in $1/N$ one deals with manifestly conformal invariant two and
three point functions whose dependence on momenta is fixed by
conformal invariance. The only unknown quantities: the scale
dimensions of $\psi$ and $\bar\psi\psi$ field as well as the critical
value of the coupling constant are determined order by order in $1/N$.
In $d=3$ we have shown by explicit calculations how the logarithms of
perturbative theory are nicely collected into conformally invariant
structures.  As is discussed above, the reason behind this is the
exponentiation of the log's into scale-invariant expressions which is
a nonperturbative phenomenon from the viewpoint on the
$1/N$-expansion.

We have demonstrated that the calculation of critical indices by the
method of conformal bootstrap is an economical one in the sense that
the number of skeleton graphs in the vertex bootstrap equation is much
less than the total number of graphs of the perturbative expansion to
the given order in $1/N$.  Moreover, one only needs to calculate the
vertex function while the two propagator equations are expressed via
it.

It would be very interesting to extend the conformal bootstrap
approach to the current-current four-Fermi interaction and to quantum
electrodynamics. In these cases one might expect further
simplifications due to the conservation of the vector current.

\section*{Acknowledgments}

We thank M. Dobroliubov for discussions.  Yu.M.\ thanks UBC Physics
Department for the hospitality at Vancouver last spring. W.C.\ and
G.W.S.\ were supported in part by the Natural Sciences and Engineering
Research Council of Canada.

\eop
\setcounter{section}{0}
\appendix{Parisi procedure at small $\gamma$}

Here we derive the conformal boostrap equations from Schwinger-Dyson
equations for the self-enegies, for small $\gamma$.
According to \rf{finally1} and \rf{finally2},
we should extract the $\epsilon^{-1}$ term in
$\Gamma^\epsilon S \Gamma_\alpha$ and $\Gamma^\epsilon S^\p \Gamma_\alpha$.
To the leading order in $\gamma$  we have
\bea
\hbox{Tr}\,{\bf [}\Gamma^\epsilon S \Gamma_\alpha {\bf ]}
&=-&\hbox{Tr}\,\int \frac{dq}{\pi^h}
\Gamma^\epsilon(q,p+q)G(p+q)\frac{\d}{\d p_\alpha}\Gamma(p+q,q)
G(q) \nonumber \\
&=& \hbox{Tr}\, \int \frac{dq}{\pi^h} \frac{g^2_*\gamma\Gamma(h)}
{\hbox{\rm max}\{q^2,p^2\}^{\gamma+\epsilon/2}}
\frac{\hat{q}}{[q^2]^{h-l+1/2}}
\frac{\hat{q}+\hat{p}}{[(q+p)^2]^{h-l+1/2}} \nonumber \\
& &\cdot\frac{dk}{\pi^h}\frac{1}{[(q-k)^2]^{l-b/2}}
\frac{\hat{k}}{[k^2]^{b/2+1/2}}\frac{\d}{\d p_\alpha }
\frac{\hat{k}-\hat{p}}{[(k-p)^2]^{b/2+1/2}}\,.
\eea
Since the integral over $dq$ equals $2\pi^h/\epsilon\Gamma(h)$, it turns out to
be
\bea
\hbox{Tr}\, {\bf [}\Gamma^\epsilon S \Gamma_\alpha {\bf ]} &= &
\frac {2}{\epsilon} g^2_* \gamma \hbox{Tr}\,\frac{\d}{\d p_\alpha }
 \int \frac{dk}{\pi^h} \frac{\hat{k}}{[k^2]^{b/2+1/2}}
\frac{\hat{k}-\hat{p}}{[(k-p)^2]^{b/2+1/2}} \nonumber \\
&=&\frac {2}{\epsilon} g^2_* \gamma \hbox{Tr}\,{\bf 1}\frac{\d}{\d
p_\alpha}
[p^2]^{h-b}N(d-b) \tilde{N}^2(\frac b2).
\label{11}
\eea

Similarly, we obtain
\be
\Gamma^\epsilon S^\p \Gamma_\alpha
=\frac 2\epsilon g^2_* \gamma
\frac{\d}{\d p_\alpha }
[i\hat{p}(p^2)^{h-l-1/2}]\tilde{N}(\frac b2)\tilde{N}(d-l) N(l-\frac b2).
\label{22}
\ee

Substituting Eqs.~\rf{11}, \rf{22} into \rf{finally1} and \rf{finally2},
we arrive at, in the small $\gamma$-limit,
\rf{ve}, \rf{fe} and
\rf{be}.

\eop

\eop

\section*{Figures}

\unitlength=1.00mm
\linethickness{0.5pt}
\thicklines
\vspace{0.5cm}

\begin{picture}(110.00,50.00)
\thicklines
\put(104.5,18.00){\circle{20.00}}
\put(90.5,18.00){\circle{20.00}}
\put(56.0,18.00){\circle{20.00}}
\put(70.0,18.00){\circle{20.00}}
\put(79.00,18.00){\makebox(0,0)[cc]{ \ $\ldots$}}
\put(49.0,18.00){\line(-4,3){4.00}}
\put(49.0,18.00){\line(-4,-3){4.00}}
\put(111.50,18.00){\line(4,3){4.00}}
\put(111.50,18.00){\line(4,-3){4.00}}

\end{picture}
\begin{description}
\item[Fig. 1]
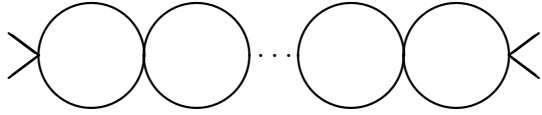
\ \ \ \ The four-Fermion scattering in the four-Fermion theories.
The solid line is the fermion propagator.
\end{description}
\vspace{2.cm}
\begin{picture}(170.00,62.00)
\thicklines
\multiput(0.00,18.)(2.00,0.00){8}{\line(3,0){1.00}}
\multiput(0.500,17.5)(2.00,0.00){8}{\line(0,3){1.00}}
\put(23.00,18.00){\makebox(0,0)[cc]{ = }}
\multiput(31.00,18)(2.00,0.00){8}{\line(3,0){1.00}}
\put(51.00,18.00){\makebox(0,0)[cc]{ + }}
\multiput(56.00,18)(2.00,0.00){4}{\line(3,0){1.00}}
\put(70.0,18.00){\circle{20.00}}
\multiput(77.00,18.0)(2.00,0.00){4}{\line(3,0){1.00}}
\put(89.00,18.00){\makebox(0,0)[cc]{ + }}
\multiput(94.00,18.0)(2.00,0.00){4}{\line(3,0){1.00}}
\put(108.0,18.00){\circle{20.00}}
\multiput(115.00,18.0)(2.00,0.00){3}{\line(3,0){1.00}}
\put(127.0,18.00){\circle{20.00}}
\multiput(134.00,18.00)(2.00,0.00){4}{\line(3,0){1.00}}
\put(149.00,18.00){\makebox(0,0)[cc]{ + \ $\ldots$}}

\end{picture}

\begin{description}
\item[Fig. 2]
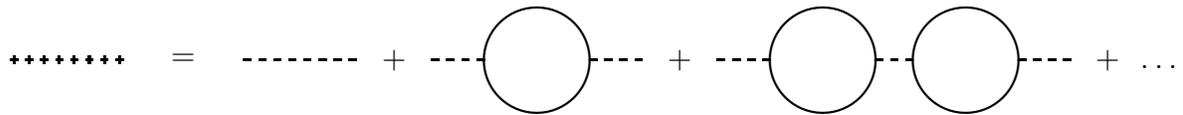
\ \ \ \ The summation of infinite series of one-loop Fermion
bubble chains gives the dressed scalar propagator, which
is of order ${\cal O}(N^0)$.
The cross line is the dressed scalar propagator, the dashed
line is the `bare' scalar propagator.
\end{description}
\unitlength=1.00mm
\linethickness{0.5pt}
\begin{picture}(160.00,80.00)
\thicklines
\put(29.25,12.50){\line(1,0){21.00}}
\multiput(33.00,14.00)(1.00,2.00){4}{\line(3,0){1.00}}
\multiput(45.00,14.00)(-1.00,2.00){4}{\line(3,0){1.00}}
\multiput(38.00,21.00)(2.00,0.00){2}{\line(3,0){1.00}}
\multiput(33.500,13.500)(1.00,2.00){4}{\line(0,3){1.00}}
\multiput(45.500,13.500)(-1.00,2.00){4}{\line(0,3){1.00}}
\multiput(38.5,20.500)(2.00,0.00){2}{\line(0,3){1.00}}
\put(39.00,4.00){\makebox(0,0)[cc]{$a$}}
\put(92.0,12.50){\line(1,0){16.00}}
\put(100.00,12.50){\makebox(0,0)[cc]{$\times$}}
\put(100.00,4.00){\makebox(0,0)[cc]{$a^\prime$}}

\end{picture}
\begin{description}
\item[Fig. 3]
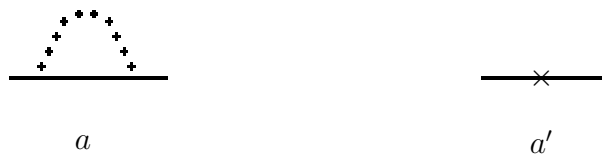
\ \ \ \ The fermion self-energy at ${\cal O}(1/N)$. $a^\prime$ is
the proper counterterm.
\end{description}
\vspace{2.0cm}
\begin{picture}(160.00,45.00)

\thicklines

\multiput(40.00,20.0)(0.00,2.00){3}{\line(0,3){1.00}}
\multiput(39.50,20.50)(0.00,2.00){3}{\line(3,0){1.00}}
\put(40.00,20.00){\line(1,-2){8.00}}
\put(40.00,20.00){\line(-1,-2){8.00}}
\multiput(34.60,8.00)(2.00,0.00){6}{\line(3,0){1.00}}
\multiput(100.00,12.0)(0.00,2.0){3}{\line(0,3){1.00}}
\multiput(35.10,7.50)(2.00,0.00){6}{\line(0,3){1.00}}
\multiput(99.50,12.50)(0.00,2.0){3}{\line(3,0){1.00}}
\put(92.0,10.50){\line(1,0){16.0}}
\put(100.055,10.625){\makebox(0,0)[cc]{$\times$}}

\put(40.00,-5.00){\makebox(0,0)[cc]{$a$}}
\put(100.00,-5.00){\makebox(0,0)[cc]{$a^\prime$}}

\end{picture}
\vspace{1cm}
\begin{description}
\item[Fig. 4]
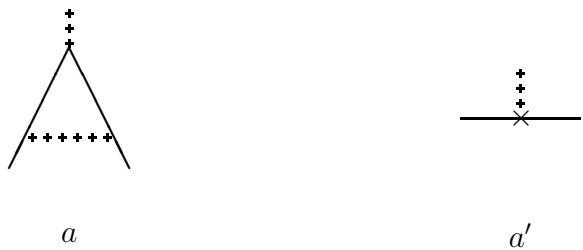
\ \ \ \ The vertex correction at order ${\cal O}(1/N)$.
$a^\prime$ is the proper counterterm.
\end{description}

\unitlength=0.90mm

\begin{picture}(110.00,75.00)

\thicklines

\put(30.0,20.00){\circle{30.00}}
\multiput(21.00,20)(-2.00,0.00){3}{\line(3,0){1.00}}
\multiput(37.00,23)(2.00,0.00){4}{\line(3,0){1.00}}
\multiput(37.00,17)(2.00,0.00){4}{\line(3,0){1.00}}
\multiput(21.50,19.5)(-2.00,0.00){3}{\line(0,3){1.00}}
\multiput(37.50,22.5)(2.00,0.00){4}{\line(0,3){1.00}}
\multiput(37.50,16.5)(2.00,0.00){4}{\line(0,3){1.00}}
\put(44,28){\line(0,-2){16}}
\put(75.0,20.00){\circle{30.00}}
\multiput(66.00,20)(-2.00,0.00){3}{\line(3,0){1.00}}
\multiput(82.50,23)(2.00,0.00){4}{\line(3,0){1.00}}
\multiput(82.50,17)(2.00,0.00){4}{\line(3,0){1.00}}
\multiput(105.00,20)(2.00,0.00){3}{\line(3,0){1.00}}
\multiput(66.50,19.5)(-2.00,0.00){3}{\line(0,3){1.00}}
\multiput(83.0,22.5)(2.00,0.00){4}{\line(0,3){1.00}}
\multiput(83.0,16.5)(2.00,0.00){4}{\line(0,3){1.00}}
\multiput(105.50,19.5)(2.00,0.00){3}{\line(0,3){1.00}}
\put(97.0,20.00){\circle{30.00}}
\put(33.00,5.0){\makebox(0,0)[cc]{$a$}}
\put(86.50,5.0){\makebox(0,0)[cc]{$b$}}

\end{picture}
\begin{description}
\item[Fig. 5]
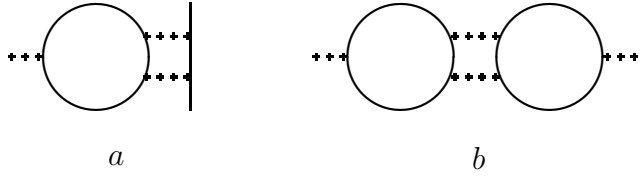
\ \ \ \ The null diagrams in the symmetric phase at order $O(1/N)$.
\end{description}

\unitlength=1.0mm

\begin{picture}(110.00,110.00)

\thicklines

\put(20.0,80.00){\circle{30.00}}
\multiput(12.00,80)(-2.00,0.00){3}{\line(3,0){1.00}}
\multiput(27.00,80)(2.00,0.00){3}{\line(3,0){1.00}}
\multiput(14.50,76)(1.00,2.00){3}{\line(3,0){1.00}}
\multiput(24.750,76)(-1.00,2.00){3}{\line(3,0){1.00}}
\multiput(18.50,81)(2.00,0.00){2}{\line(3,0){1.00}}
\multiput(12.50,79.5)(-2.00,0.00){3}{\line(0,3){1.00}}
\multiput(27.50,79.5)(2.00,0.00){3}{\line(0,3){1.00}}
\multiput(15.0,75.5)(1.00,2.00){3}{\line(0,3){1.00}}
\multiput(25.250,75.5)(-1.00,2.00){3}{\line(0,3){1.00}}
\multiput(19.0,80.5)(2.00,0.00){2}{\line(0,3){1.00}}
\put(70.0,80.00){\circle{30.00}}
\multiput(62.00,80)(-2.00,0.00){3}{\line(3,0){1.00}}
\multiput(77.0,80)(2.00,0.00){3}{\line(3,0){1.00}}
\multiput(62.50,79.5)(-2.00,0.00){3}{\line(0,3){1.00}}
\multiput(77.5,79.5)(2.00,0.00){3}{\line(0,3){1.00}}
\put(70.00,73.0){\makebox(0,0)[cc]{$\times$}}
\put(20.00,65.0){\makebox(0,0)[cc]{$a$}}
\put(70.00,65.0){\makebox(0,0)[cc]{$a^\prime$}}
\put(20.0,50.00){\circle{30.00}}
\multiput(12.00,50)(-2.00,0.00){3}{\line(3,0){1.00}}
\multiput(27.00,50)(2.00,0.00){3}{\line(3,0){1.00}}
\multiput(12.50,49.5)(-2.00,0.00){3}{\line(0,3){1.00}}
\multiput(27.50,49.5)(2.00,0.00){3}{\line(0,3){1.00}}
\multiput(14.50,54.)(1.00,-2.00){3}{\line(3,0){1.00}}
\multiput(24.750,54.)(-1.00,-2.00){3}{\line(3,0){1.00}}
\multiput(18.650,49.1)(2.00,0.00){2}{\line(3,0){1.00}}
\multiput(15.0,53.5)(1.00,-2.00){3}{\line(0,3){1.00}}
\multiput(25.250,53.5)(-1.00,-2.00){3}{\line(0,3){1.00}}
\multiput(19.150,48.6)(2.00,0.00){2}{\line(0,3){1.00}}
\put(70.0,50.00){\circle{30.00}}
\multiput(62.00,50)(-2.00,0.00){3}{\line(3,0){1.00}}
\multiput(77.0,50)(2.00,0.00){3}{\line(3,0){1.00}}
\multiput(62.50,49.5)(-2.00,0.00){3}{\line(0,3){1.00}}
\multiput(77.5,49.5)(2.00,0.00){3}{\line(0,3){1.00}}
\put(70.00,57.0){\makebox(0,0)[cc]{$\times$}}
\put(20.00,35.0){\makebox(0,0)[cc]{$b$}}
\put(70.00,35.0){\makebox(0,0)[cc]{$b^\prime$}}
\put(20.0,20.00){\circle{30.00}}
\multiput(12.00,20)(-2.00,0.00){3}{\line(3,0){1.00}}
\multiput(27.00,20)(2.00,0.00){3}{\line(3,0){1.00}}
\multiput(12.50,19.5)(-2.00,0.00){3}{\line(0,3){1.00}}
\multiput(27.50,19.5)(2.00,0.00){3}{\line(0,3){1.00}}
\multiput(19.50,26.0)(0.00,-2.00){7}{\line(3,0){1.00}}
\multiput(20.00,25.50)(0.00,-2.00){7}{\line(0,3){1.00}}
\put(70.0,20.00){\circle{30.00}}
\multiput(62.00,20)(-2.00,0.00){3}{\line(3,0){1.00}}
\multiput(77.0,20)(2.00,0.00){3}{\line(3,0){1.00}}
\multiput(62.50,19.5)(-2.00,0.00){3}{\line(0,3){1.00}}
\multiput(77.5,19.5)(2.00,0.00){3}{\line(0,3){1.00}}
\put(63.00,20.0){\makebox(0,0)[cc]{$\times$}}
\put(120.0,20.00){\circle{30.00}}
\multiput(112.00,20)(-2.00,0.00){3}{\line(3,0){1.00}}
\multiput(127.00,20)(2.00,0.00){3}{\line(3,0){1.00}}
\multiput(112.50,19.5)(-2.00,0.00){3}{\line(0,3){1.00}}
\multiput(127.50,19.5)(2.00,0.00){3}{\line(0,3){1.00}}
\put(127.00,20.0){\makebox(0,0)[cc]{$\times$}}
\put(20.00,5.0){\makebox(0,0)[cc]{$c$}}
\put(70.00,5.0){\makebox(0,0)[cc]{$c^\prime$}}
\put(120.00,5.0){\makebox(0,0)[cc]{$c^\prime$$^\prime$}}
\end{picture}
\begin{description}
\item[Fig. 6]
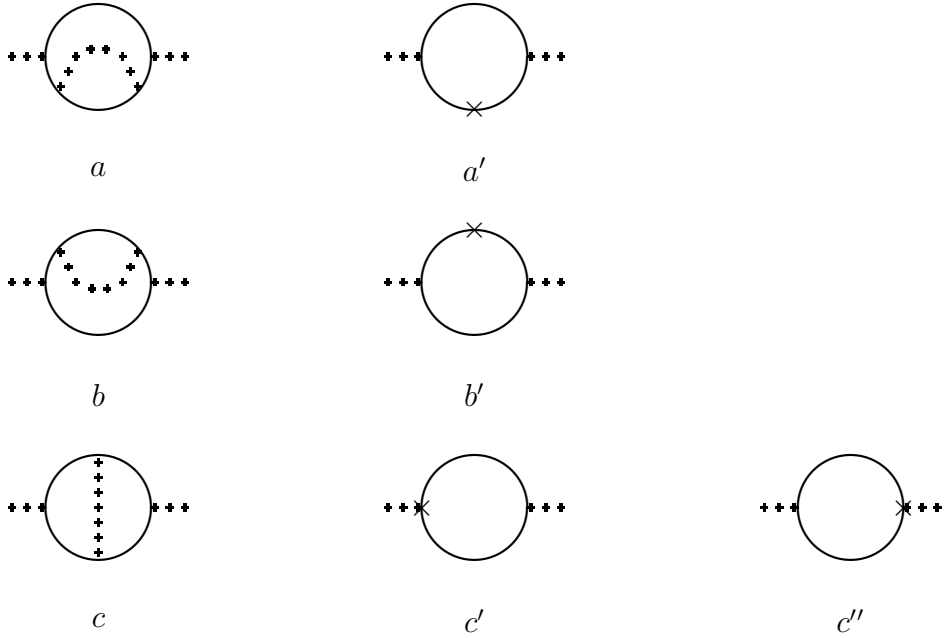
\ \ \ \ The scalar self energy at order ${\cal O}(1/N)$.
$a^\prime$, $b^\prime$, $c^\prime$ and $c^\prime$$^\prime$ are due to the
proper counterterms.
\end{description}

\vspace{6cm}
\thinlines

\begin{picture}(92.00,50.00)
\linethickness{0.2pt}

\put(80.00,24.00){\circle*{2.00}}
\put(80.20,24.0){\line(-1,-2){4.00}}
\put(80.20,24.0){\line(1,-2){4.00}}
\multiput(80.20,23.0)(0.00,2.00){5}{\line(0,3){1.00}}
\put(79.80,24.0){\line(-1,-2){4.00}}
\put(79.80,24.0){\line(1,-2){4.00}}
\multiput(79.80,23.0)(0.00,2.00){5}{\line(0,3){1.00}}
\put(60.00,20.00){\makebox(0,0)[cc]{$\Gamma$ \ \ \ \ $=$}}
\end{picture}

\begin{description}
\item[Fig. 7]\ \ \ The conformal Yukawa vertex  constructed by truncating
the external legs of the conformal
three-point function $\langle\phi\psi\bar\psi\rangle$.
\end{description}

\vspace{5cm}

\begin{picture}(150.00,38.41)
\linethickness{0.2pt}

\put(10.00,20.00){\makebox(0,0)[cc]{$K$ \ \ \ $=$}}
\put(40.00,10.00){\circle*{2.00}}
\put(40.00,30.00){\circle*{2.00}}
\put(20.0,10.20){\line(1,0){40.00}}
\multiput(40.20,10.50)(0.00,2.00){10}{\line(0,3){1.00}}
\put(20.0,30.20){\line(1,0){40.00}}
\put(20.0,9.80){\line(1,0){40.00}}
\multiput(39.80,10.50)(0.00,2.00){10}{\line(0,3){1.00}}
\put(20.0,29.8){\line(1,0){40.00}}
\put(70.00,20.0){\makebox(0,0)[cc]{$+$ \ $\ldots\;,$}}
\put(40.00,0.00){\makebox(0,0)[cc]{$a$}}
\put(120.00,10.00){\circle*{2.00}}
\put(120.00,10.20){\line(1,0){20.00}}
\multiput(100.0,10.20)(2.00,0.00){10}{\line(1,0){1.00}}
\put(120.2,10){\line(0,1){20}}
\put(100.00,30.20){\line(1,0){20.00}}
\multiput(121.00,30.20)(2.00,0.00){10}{\line(1,0){1.00}}
\put(120.00,9.80){\line(1,0){20.00}}
\multiput(100.00,9.80)(2.00,0.00){10}{\line(1,0){1.00}}
\put(119.8,10){\line(0,1){20}}
\put(100.00,29.80){\line(1,0){20.00}}
\multiput(121.00,29.80)(2.00,0.00){10}{\line(1,0){1.00}}
\put(120.00,30.00){\circle*{2.00}}
\put(90.00,20.00){\makebox(0,0)[cc]{$K^\prime$ \ \ \ $=$}}
\put(150.00,20.00){\makebox(0,0)[cc]{$+$ \ $\ldots$}}
\put(120.00,0.00){\makebox(0,0)[cc]{$b$}}
\end{picture}

\vspace{1cm}
\begin{description}
\item[Fig. 8]
\ \ \ \ The skeleton expansions of $a)$ Fermion-Fermion
 and $b)$ boson-Fermion Bethe--Salpeter kernel.
The double solid line is the full Fermion two-point function,
the double dashed line is the full scalar two-point
function.
\end{description}

\begin{picture}(90.00,80.00)
\linethickness{0.2pt}

\put(40.00,17.00){\circle*{2.00}}
\put(40.20,17.0){\line(-1,-2){4.00}}
\put(40.20,17.0){\line(1,-2){4.00}}
\multiput(40.2,16.0)(0.00,2.00){5}{\line(0,3){1.00}}
\put(39.80,17.0){\line(-1,-2){4.00}}
\put(39.80,17.0){\line(1,-2){4.00}}
\multiput(39.80,16.0)(0.00,2.00){5}{\line(0,3){1.00}}
\put(56.00,17.00){\makebox(0,0)[cc]{$=$}}
\put(80.00,30.00){\circle*{2.00}}
\put(70.50,11.00){\circle*{2.00}}
\put(89.50,11.00){\circle*{2.00}}
\multiput(80.20,29.0)(0.00,2.00){5}{\line(0,3){1.00}}
\put(80.20,30.0){\line(-1,-2){13.00}}
\put(80.20,30.0){\line(1,-2){13.00}}
\multiput(70.50,11.2)(2.00,0.00){10}{\line(3,0){1.00}}
\multiput(79.80,29.0)(0.00,2.00){5}{\line(0,3){1.00}}
\put(79.80,30.0){\line(-1,-2){13.00}}
\put(79.80,30.0){\line(1,-2){13.00}}
\multiput(70.50,10.80)(2.00,0.00){10}{\line(3,0){1.00}}
\put(67.00,22.00){\makebox(0,0)[cc]{$k+p_1$}}
\put(80.00,7.80){\makebox(0,0)[cc]{$k$}}
\put(93.00,22.00){\makebox(0,0)[cc]{$k+p_2$}}

\end{picture}
\vspace{1cm}
\begin{description}
\item[Fig. 9]
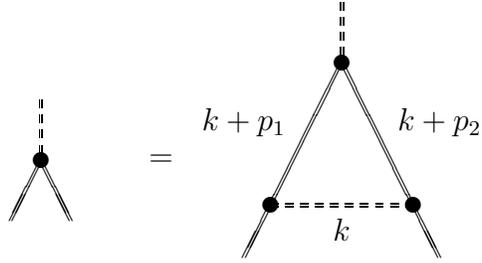
\ \ \ \ The three-vertex approximation of the vertex bootstrap equation.
\end{description}

\end{document}